# Asymmetric Electromagnetic Modes on a Two-Dimensional InSb-Air-InSb Waveguide

Saeed Pahlavan
Electronic and Computer Engineering Faculty
Tarbiat Modares University
Tehran, Iran
s_pahlavan@modares.ac.ir

*Abstract*— In this paper, the asymmetric propagation of electromagnetic waves inside a two-dimensional air slit cut in an InSb slab is studied. It has been shown that due to the anisotropic magnetic properties of InSb under a DC magnetic bias, forward and backward waves show different field patterns inside the air slit. The proposed waveguide can have potential asymmetric applications in mmWave and sub-THz regions.

*Keywords—Magneto-optic; Anisotropic; Propagation; Dispersion; Permittivity tensor; Chiral.*

## I. INTRODUCTION

Magnetic materials are subject to numerous studies and promising applications such as microwave isolators and filters are introduced to the market based on the anisotropic permeability of these materials. The anisotropy shows itself in terms of a permeability tensor with unequal off-diagonal elements based on the direction of the external magnetic bias. For example, if the field is applied along the z direction the permeability is [1]:

$$\bar{\bar{\mu}} = \mu_0 \begin{bmatrix} \mu_1 & -j\mu_2 & 0 \\ j\mu_2 & \mu_1 & 0 \\ 0 & 0 & \mu_3 \end{bmatrix} \quad (1)$$

Solving wave equations in structures containing magnetic materials unveils some unique behaviors that can never occur in ordinary microstrip, rectangular or parallel plate waveguides, or even periodic structures, etc. Hence if a metal-dielectric-metal structure is reciprocal, including magnetic materials such as ferrites can make the structure behave as an isolator [2-8].

However, magnetic materials only show this interesting property in low microwave frequencies, while the need for isolators in mmWave, terahertz, infrared, and optical frequencies has been increasing in the past decade due to the increasing need for higher bandwidths and smaller circuit sizes. This encouraged us to seek new solutions at these frequencies.

Among all chiral materials [9], magneto-optic materials show an interesting property under an external magnetic bias field. In the sub-terahertz region as an example, InSb shows a permittivity tensor that depends on the direction of the external field.

In this paper, we analyze the electromagnetic properties of a two-dimensional air-filled waveguide surrounded by InSb as a magneto-optic material and discuss the differences between the proposed structure and a similar metallic structure [10].

## II. NEGATIVE PERMITTIVITY IN MAGNETO OPTIC MATERIALS

The electric permittivity of a magneto-optic material behaves like the magnetic permeability of a ferrite. However, their operating frequencies are different. While ferrites work at microwave, InSb is suitable for near terahertz regions since under a z-directed DC bias field $\vec{B_0}$ the permittivity tensor will be [9]:

$$\bar{\bar{\epsilon}} = \epsilon_0 \begin{bmatrix} \epsilon_1 & -j\epsilon_2 & 0 \\ j\epsilon_2 & \epsilon_1 & 0 \\ 0 & 0 & \epsilon_3 \end{bmatrix} \quad (2.1)$$

$$\epsilon_1 = \epsilon_\infty - \frac{\omega_p^2(\omega^2 + i\gamma\omega)}{(\omega^2 + i\gamma\omega)^2 - \omega^2\omega_c^2} \quad (2.2)$$

$$\epsilon_2 = \frac{\omega\omega_c\omega_p^2}{(\omega^2 + i\gamma\omega)^2 - \omega^2\omega_c^2} \quad (2.3)$$

$$\epsilon_3 = \epsilon_\infty - \frac{\omega_p^2}{(\omega^2 + i\gamma\omega)^2} \quad (2.4)$$

here $\omega_p$ and $\omega_c$ are the plasma and cyclotron angular frequencies respectively, $\frac{\gamma}{2\pi}$ is the collision frequency, $B_0 \sim \frac{\omega_c}{\omega_p}$ is the applied magnetic flux, and $\epsilon_\infty \simeq 15.68$ is the high frequency background dielectric constant for Indium-Antimonide (InSb) [9].

The effective permittivity $\epsilon_\perp = \epsilon_1 - {\epsilon_2^2}/{\epsilon_1}$ is defined in electromagnetic field propagation for transverse magnetic (TM) waves inside a magneto-optic material [10]. It is shown in Fig. 1 that the real part of this quantity is highly negative in frequencies below 300GHz, making this material a suitable substitution for metals. In the next section, we show how this property affects wave propagation in an InSb-Air-InSb slot.

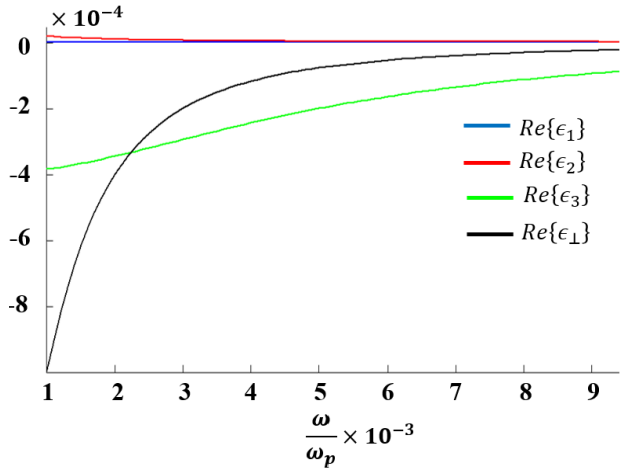

Fig. 1: Real parts of permittivity tensor elements of InSb versus normalized frequency.

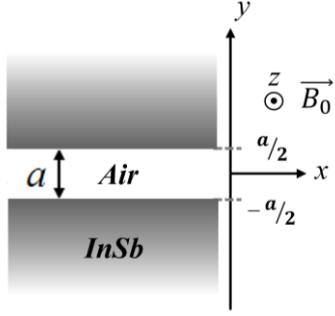

Fig. 2: Two dimensional InSb-air-InSb waveguide.

## III. Two dimensional InSb-Air-InSb waveguide

The proposed structure of Fig. 2 is a two-dimensional waveguide in the sense that it is invariant in the z-direction. Hence, it is very similar to classic parallel-plate waveguides [5] for which the Maxwell's equations with tensor permittivity can be written as:

$$\nabla \times \mathbf{E} = -j\omega\mu\mathbf{H} \Rightarrow \begin{cases} \frac{\partial E_y}{\partial z} - \frac{\partial E_z}{\partial y} = -j\omega\mu H_x \\ \frac{\partial E_z}{\partial x} - \frac{\partial E_x}{\partial z} = -j\omega\mu H_y \\ \frac{\partial E_x}{\partial y} - \frac{\partial E_y}{\partial x} = -j\omega\mu H_z \end{cases} \quad (3)$$

$$\nabla \times \mathbf{H} = j\omega\bar{\bar{\epsilon}}\mathbf{E} \Rightarrow \begin{cases} \frac{\partial H_y}{\partial z} - \frac{\partial H_z}{\partial y} = j\omega\epsilon_0(\epsilon_1 E_x - j\epsilon_2 E_y) \\ \frac{\partial H_z}{\partial x} - \frac{\partial H_x}{\partial z} = j\omega\epsilon_0(j\epsilon_2 E_x + \epsilon_2 E_y) \\ \frac{\partial H_x}{\partial y} - \frac{\partial H_y}{\partial x} = j\omega\epsilon_0\epsilon_3 E_z \end{cases} \quad (4)$$

Since $\frac{\partial}{\partial z} = 0$, Maxwell's equations result in two separate Helmholtz's equations; a transverse electric field and a transverse magnetic field:

$$TE: \nabla_{x,y}^2 E_z + k_0^2 \epsilon_3 E_z = 0 \quad (5.1)$$

$$TM: \nabla_{x,y}^2 H_z + k_0^2 \epsilon_\perp H_z = 0 \quad (5.2)$$

In the frequency range that $\epsilon_\perp$ is negative, the TM wave components have the general form of:

$$H_z^{air} = (A\sinh\alpha_0 y + B\cosh\alpha_0 y)e^{-j\beta x} \quad (6.1)$$

$$E_x^{air} = \frac{1}{j\omega\epsilon_0}\alpha_0 e^{-j\beta x}(B\sinh\alpha_0 y + A\cosh\alpha_0 y) \quad (6.2)$$

and

$$H_z^{MO} = e^{-j\beta x}\begin{cases} Ce^{-\alpha(y-\frac{a}{2})} & y>a/2 \\ De^{\alpha(y+\frac{a}{2})} & y<-a/2 \end{cases} \quad (7.1)$$

$$E_x^{MO} = \begin{cases} \frac{D(+\epsilon_1\alpha-\epsilon_2\beta)}{j\omega\epsilon_0\epsilon_1\epsilon_\perp}e^{-j\beta x}e^{+\alpha(y-\frac{a}{2})} & y<\frac{-a}{2} \\ \frac{C(-\epsilon_1\alpha-\epsilon_2\beta)}{j\omega\epsilon_0\epsilon_1\epsilon_\perp}e^{-j\beta x}e^{-\alpha(y-\frac{a}{2})} & y>\frac{a}{2} \end{cases} \quad (7.2)$$

$$\alpha_0 = \sqrt{\beta^2 - k_0^2}, \quad \alpha = \sqrt{\beta^2 - k_0^2\epsilon_\perp} \quad (7.3)$$

Here, $\beta$ and $k_0$ are wave propagation constants along the x-axis inside the waveguide and in free space respectively, while $\alpha$ and $\alpha_0$ are the wave numbers along the y-axis for InSb and air respectively. Also, A, B, C, and D are unknown amplitudes that can be found by solving the governing boundary conditions. Moreover, solving the boundary conditions for tangential filed components results in the dispersion relation:

$$\left(\frac{2\sqrt{\beta^2-k_0^2\epsilon_\perp}}{\epsilon_\perp\sqrt{\beta^2-k_0^2}}\right)\coth\left(a\sqrt{\beta^2-k_0^2}\right) + 1 +$$
$$\frac{(\epsilon_1^2(\beta^2-k_0^2\epsilon_\perp)-\epsilon_2^2\beta^2)}{(\sqrt{\beta^2-k_0^2\epsilon_1\epsilon_\perp})^2} = 0 \quad (8)$$

Figure 3 shows that the propagation constant $\beta$ inside the proposed waveguide is different from the free space wave constant. Needless to mention that for a PCE parallel plate waveguide, the fundamental mode is TEM propagating with $k_0$.

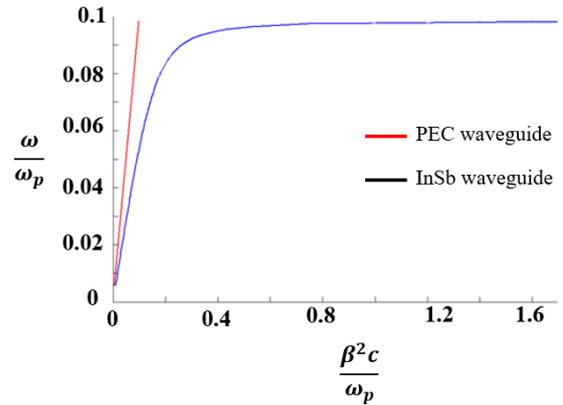

Fig. 3: Normalized propagation constant inside the air slit; comparison between PEC and InSb waveguide.

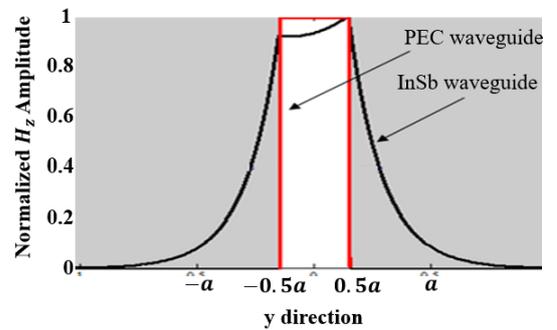

Fig. 4: Normalized transverse magnetic field amplitude; comparison between PEC and InSb waveguide.

This asymmetry is obvious in normalized transverse magnetic field patterns based on equations (7) and (8) as plotted in Fig. 4. Obviously, based on (7.2) the electric field is not equal for forward and backward waves since it's not an even function of $\beta$. Removing the external bias turns the InSb into a homogeneous medium due to a vanishing $\varepsilon_2$. Under such conditions, changing the sign of $\beta$ does not change the electric field patterns.

### IV. Conclusion

A novel two-dimensional waveguide with asymmetric forward and backward propagating modes based on the asymmetric permittivity tensor of InSb as the waveguide constituent was presented and studied. The proposed waveguide is an air-slit cut inside an InSb slab under an external DC magnetic field bias. The propagating field patterns derived by the mode-matching technique show different patterns in different propagation directions which is promising for mmWave and sub-THz asymmetric applications.